\documentclass[showpacs,preprintnumbers,superscriptaddress,prb,floatfix,aps,10pt,preprint]{revtex4-1}

\usepackage{hyperref}
\hypersetup{
    bookmarks=true,			
    unicode=true,           	
    pdftoolbar=false,       
    pdfmenubar=true,        
    pdffitwindow=true,      
    pdftitle={The origin of the low kappa},		
    pdfauthor={Nina Shulumba},					
    pdfsubject={Subject},   
    pdfnewwindow=true,      
    pdfkeywords={keywords}, 
    colorlinks=true,		
    linkcolor=blue,         
    citecolor=blue,         
    filecolor=black,      	
    urlcolor=blue           
			}

\usepackage{amsmath,amssymb}
\usepackage{graphicx,textcomp}
\usepackage{subfigure}
\usepackage{color}
\usepackage{mhchem}
\usepackage{soul,xcolor}
\setstcolor{red}

\usepackage[titletoc,toc,title]{appendix}

\usepackage{todonotes}
\usepackage[normalem]{ulem}
\usepackage[titletoc,toc,title]{appendix}

\usepackage{xcolor}

\usepackage[title,titletoc,toc]{appendix}
\newcommand{\figref}[1]{Fig.~\ref{#1}}
\newcommand{\figrefmany}[1]{Figs.~\ref{#1}}
\newcommand{\figreffull}[1]{Figure~\ref{#1}}
\newcommand{\secref}[1]{Sec.~\ref{#1}}

\renewcommand {\vec}    [1]    {\ensuremath{\mathbf{#1}}}

\newcommand	  {\set}    [1]    {\ensuremath{ \{ #1 \} }}

\begin{document}

\title{Intrinsic localized mode and low thermal conductivity of PbSe}

\author{Nina Shulumba}
\affiliation{Division of Engineering and Applied Science, California Institute of Technology, Pasadena, California 91125, USA}
\author{Olle Hellman}
\affiliation{Division of Engineering and Applied Science, California Institute of Technology, Pasadena, California 91125, USA}
\affiliation{Link\"{o}ping University, Department of Physics, Chemistry, and Biology (IFM), Link\"oping SE-581 83, Sweden.}
\author{Austin J. Minnich}
\affiliation{Division of Engineering and Applied Science, California Institute of Technology, Pasadena, California 91125, USA}

\date{\today}

\begin{abstract}

Lead chalcogenides such as PbS, PbSe, and PbTe are of interest for their exceptional thermoelectric properties and strongly anharmonic lattice dynamics. Although PbTe has received the most attention, PbSe has a lower thermal conductivity despite being stiffer, a trend that prior first-principles calculations have not reproduced. Here, we use \emph{ab-initio} calculations that explicitly account for strong anharmonicity to identify the origin of this low thermal conductivity as an anomalously large anharmonic interaction, exceeding in strength that in PbTe, between the transverse optic and longitudinal acoustic branches. The strong anharmonicity is reflected in the striking observation of an intrinsic localized mode that forms in the acoustic frequencies. Our work shows the deep insights into thermal phonons that can be obtained from \emph{ab-initio} calculations that are not confined to the weak limit of anharmonicity.
\end{abstract}

\maketitle

\section{Introduction}

Lead chalcogenides have been studied for decades due to their superior thermoelectric performance and strongly anharmonic lattice dynamics \cite{Cowley1968,Cochran1966,ravich1970semiconducting,pechmann1990electron}. For applications, these materials are established as champion thermoelectrics with high performance stemming from the intrinsically low thermal conductivity and favorable electronic structure \cite{wood_1988,nolas2001thermoelectrics,Pei2011,Biswas2012}. Scientifically, they are of interest because they fail to follow the conventional phonon picture in which anharmonicity is treated as a weak perturbation. In particular, in PbTe the strong anharmonic interaction between the transverse optical and longitudinal acoustic branches has been shown to result in an avoided crossing that has been observed using inelastic neutron scattering measurements \cite{Delaire2011}. Other measurements have uncovered an unusual double peak in the spectral function \cite{Burkhard1977,li2014,jensen_2012} and softening of the transverse optical mode near the ferroelectric transition \cite{bate_1970,alperin1972,Jantsch1983,Jiang2016} that again reflects the strong anharmonicity. These strong anharmonic interactions lead to low thermal conductivity and large Gr\"uneisen parameter~\cite{Romero2015}. PbTe has been studied with several \emph{ab-initio} approaches \cite{rabe_1985,an2008,zhang2009,tian_2012,Skelton2014,Lee2014}, which have provided important insights but have been unable to predict key features such as stiffening of transverse optical mode and nonlinear thermal resistivity with temperature increase. 

PbS and PbSe have also received attention as they possess similar electronic \cite{kohn1973,wei1997,Chen2013} and thermal properties \cite{parker_2013,Ong2013} to those of PbTe. However, the thermal conductivity of PbSe exhibits an unusual anomaly. Considering the typical metrics of thermal conductivity such as atomic masses, cutoff phonon frequencies, and acoustic-optical gap, one would expect that PbS should possess the highest thermal conductivity while PbTe should have the lowest. Indeed, prior \emph{ab-initio} calculations that use the ground state phonon dispersion predict this trend \cite{tian_2012,Lee2014,Skelton2014}. However, experimentally the lowest thermal conductivity is achieved by PbSe, followed by PbTe and then PbS with the highest value. That previous \emph{ab-initio} studies fail to predict the correct trend suggest that a key element is missing in the conventional approach to calculate thermal conductivity from first principles.
 
In this paper we use the temperature dependent effective potential method~\cite{tdep_hellman2011,tdep_hellman2013,tdep_hellman2013_3difc} (TDEP) to study the lattice dynamics of PbS, PbSe and PbTe. TDEP identifies effective force constants that best describe the potential surface at a given temperature and thus does not assume the 0 K dispersion, as was the case in prior studies. We show that TDEP can successfully reproduce  both the absolute values and trends of the thermal conductivity of each of these highly anharmonic compounds. Importantly, our calculations identify the origin of the low thermal conductivity of PbSe as an exceptionally strong anharmonic interaction that is reflected in the formation of an intrinsic localized mode in the acoustic frequencies. This work shows the powerful insights that can be obtained from first principles calculations that are not restricted to the limit of weak anharmonicity.

\section{Methods}\label{sec:methods}

We employ TDEP to calculate the thermal properties of PbS, PbSe, and PbTe. Traditional \emph{ab-initio} approaches to calculate thermal conductivity use the finite difference supercell approach \cite{esfarjani2008} or density-functional perturbation theory \cite{baroni_2001,bte_broido2007} to determine harmonic and anharmonic force constants at 0 K. In the former method, forces are recorded as atoms are sequentially perturbed from their equilibrium locations at 0 K. In the latter method, the force constants are determined from the analytical derivatives of the potential energy. Although these methods have been very successful for a large number of crystals \cite{lindsay2013,nicole_2014,mingo2014,mingowu_2014,Alan2016}, they are not suitable for solids that exhibit substantial changes in phonon dispersions with temperature; in other words, for highly anharmonic solids. In the present case, lead chalcogenides are well-known to exhibit both softening and stiffening with temperature, depending on the particular mode, and as a consequence the traditional \emph{ab-initio} approach is unable to reproduce key features of the phonon dispersion and thermal conductivity temperature dependence \cite{bte_hellman,Romero2015}.

In TDEP, rather than calculating the force constants based on the equilibrium structure at 0 K, we sample the Born-Oppenheimer (BO) surface of a supercell at a given temperature and map it to a model potential energy of the following form:
\begin{equation}\label{eq:hamiltonian}
U= U_0+\frac{1}{2!}\sum_{ij\alpha\beta}\Phi_{ij}^{\alpha\beta}
u_i^\alpha u_j^\beta
+\frac{1}{3!}\sum_{ijk\alpha\beta\gamma}\Phi_{ijk}^{\alpha\beta\gamma}
u_i^\alpha u_j^\beta u_k^\gamma\,,
\end{equation}
where $u_i$ is the displacement of atom $i$ and $\alpha \beta \gamma$ are Cartesian components, and $\Phi$ are the second and third order effective interatomic force constants (IFCs). The IFCs are denoted as effective since they are identified  as the force constants that best describe the potential surface at each temperature. $U_0$ is the reference energy of the model system defined for each temperature.

A TDEP calculation consists of thermostatting a supercell and subsequently recording the forces and displacements versus time. The forces are calculated from first-principles. The force constants that best explain these force-displacement datasets are then obtained with a least-squares algorithm. In the previous papers \textit{ab initio} molecular dynamics was used to sample of the BO surface \cite{tdep_hellman2011,tdep_hellman2013,tdep_hellman2013_3difc,bte_hellman,Budai2014,Romero2015}.

In this work, we instead use an efficient stochastic sampling approach to prepare a simulation cell in uncorrelated thermally excited states \cite{West2006}. These snapshots can be created independently from each other and directly yield the necessary force-displacement datasets. To implement this stochastic sampling, for a cell of $N_a$ atoms with mass $m_i$ we use a harmonic normal mode transformation to generate positions $\set{u_i}$ and velocities $\set{\dot{u}_i}$ consistent with a canonical ensemble. The appropriate distribution of atomic positions and velocities are given by,
\begin{align}
u_i & = \sum_{s=1}^{3N_a} 
\epsilon_{is} \langle A_{is} \rangle \sqrt{-2\ln \xi_1}\sin 2\pi\xi_2 \\
\dot{u}_i & = \sum_{s=1}^{3N_a}
\omega_s \epsilon_{is} \langle A_{is} \rangle \sqrt{-2\ln \xi_1}\cos 2\pi\xi_2\,,
\end{align}
where $\omega^2_s$ and $\epsilon_{is}$ are eigenvalues and eigenvectors corresponding to mode $s$; $\xi_n$ represent uniform random variables between (0,1) producing the Box-Muller transform to normally distributed random numbers and $\langle A_{is} \rangle$ are the thermal average of the normal mode amplitudes~\cite{West2006}:
\begin{equation}
\langle A_{is} \rangle = 
\sqrt{\frac{\hbar (2n_s+1) }{2 m_i \omega_s}} 
\approx
\frac{1}{\omega_s}\sqrt{\frac{k_BT}{m_i}}\,,
\end{equation}
where $\hbar \omega \ll k_BT $ denotes the classical limit and the approximate amplitudes are valid. The classical limit has previously been used by \citet{West2006} and \citet{Petros2008} and the non-approximate distribution by \citet{Errea2014}, among others. 

Seeding the calculations to generate the first set of displacements requires the harmonic force constants, which are of course not available because they are the quantity to be calculated. Prior work has obtained the force constants using conventional density-functional perturbation theory phonon calculations or Born-Oppenheimer molecular dynamics, a tedious and expensive calculation. Here, we overcome this limitation in the following manner. Consider a pair potential $U(r)$ with two simple requirements:
\begin{equation}\label{eq:pair_pot}
\begin{split}
\frac{ \partial U(r)}{\partial r_{ij} }=&0\\
\frac{ \partial^2 U(r)}{\partial r_{ij}^2 }=&\frac{\eta}{r_{ij}^4}\,,
\end{split}
\end{equation}
that is, the pair potential has zero first derivative at pair distances ($r_{ij}$) of the equilibrium crystal and positive second derivatives that decay quickly with distance. These requirements force the crystal to be stable in this configuration. The IFCs can be calculated analytically from pair potentials, and in this case are given by,
\begin{equation}\label{eq:debye1}
\mathbf{\Phi}_{ij}(\mathbf{r})=
-\frac{\eta}{r^6}
\begin{pmatrix}
r_x^2 & r_x r_y & r_x r_z \\
r_x r_y & r_y^2 & r_y r_z \\
r_x r_z & r_y r_z & r_z^2  
\end{pmatrix}\,,
\end{equation}
where $\mathbf{r}$ is the vector between atom $i$ and $j$. This procedure gives a set of IFCs and thus a normal mode transformation that depends on a single parameter $\eta$. This parameter is determined by numerically matching the zero-point energy of the phonons to a Debye model,
\begin{equation}\label{eq:debye2}
\frac{1}{N_a} \sum_i \frac{\hbar \omega_i(\eta)}{2} = \frac{9 k_B T_D}{8} \,.
\end{equation}
Using Eqs. \eqref{eq:debye1} and \eqref{eq:debye2} we obtain a set of force constants defined by a Debye temperature. These phonons have the symmetry of the original crystal by construction and span the correct frequency range and can thus be used to seed stochastic calculations. The initial seed is used to calculate new IFCs by fitting the force-displacement dataset with the model potential energy of Eq.\eqref{eq:hamiltonian}, that in turn are used to generate new stochastic configurations until convergence. This approach allows us to create uncorrelated snapshots of the system with thermal displacements at a given temperature and volume without requiring \textit{ab initio} molecular dynamics.

Once the snapshots are created, we perform a series of first-principles simulations to obtain a set of force-displacement data sets. In this work we use with the projector augmented wave (PAW) method \cite{vasp_paw_blochl1994} as implemented in the Vienna \textit{Ab initio} Simulation Package (VASP) \cite{vasp_kresse1993,vasp_kresse1996,vasp_kresse1996_pwbs,vasp_kresse1999}. Exchange-correlation was treated using the AM05 functional \cite{am05_1,am05_2}, and the plane wave energy cutoff was set to 600 eV. We perform calculations on a temperature-volume grid consisting of 5 temperatures and 5 volumes sampled using $3\times 3\times 3$ Monkhorst-Pack~\cite{vasp_kpoints} mesh of k-points. We calculated forces and displacements from 150 configurations for each temperature and volume to ensure sufficient constraints to the IFCs. Then, we minimize Helmholtz free energy $F(T,V)$ at each temperature to find the equilibrium volume at each temperature.

We employed a $5\times 5\times 5$ (250 atom) supercell for each compound. We found that the phonon dispersions are extremely sensitive to the finite size effects. For the harmonic and cubic IFCs we truncated the force constant cutoffs at 11 and 6 coordination shells, respectively, to ensure the convergence of the phonon spectra and thermal conductivity. The detailed procedure for extracting the second and the third order IFCs from the set of forces and displacements while including the symmetry constraints has been described in Refs. \cite{tdep_hellman2011,tdep_hellman2013,tdep_hellman2013_3difc}.

The thermal conductivity is calculated by solving the full Boltzmann transport equation (BTE) using an iterative method \cite{bte_broido2007} on a $35\times 35 \times 35$ q-point grid on which the momentum conservation is exactly fulfilled. For the energy conservation we employed  the tetrahedron approach~\cite{lehman1972}.~Thermal conductivity was converged with respect to q-grid density within 0.01\%. Anharmonic phonon-phonon interactions along with isotopic scattering~\cite{Tamura_1983} from the natural distribution are included in the iterative solution of BTE. We obtain the diagonal components of the thermal conductivity tensor as,
\begin{equation}
\kappa_{\alpha\alpha}=\frac{1}{V} \sum_{\vec{q}s} C_{\vec{q}s} v^2_{\alpha \vec{q}s} \tau_{\alpha \vec{q}s}\,,
\end{equation}
where $v_{\alpha\vec{q}s}$ and $\tau_{\alpha\vec{q}s}$ are the phonon group velocity and phonon lifetime of mode $\vec{q}s$ along $\alpha$ direction, respectively. $C_{\vec{q}s}$ is the specific heat per mode. 

\section{Results}

\subsection{Thermal conductivity}
\begin{figure}
\includegraphics[width=.5\linewidth]{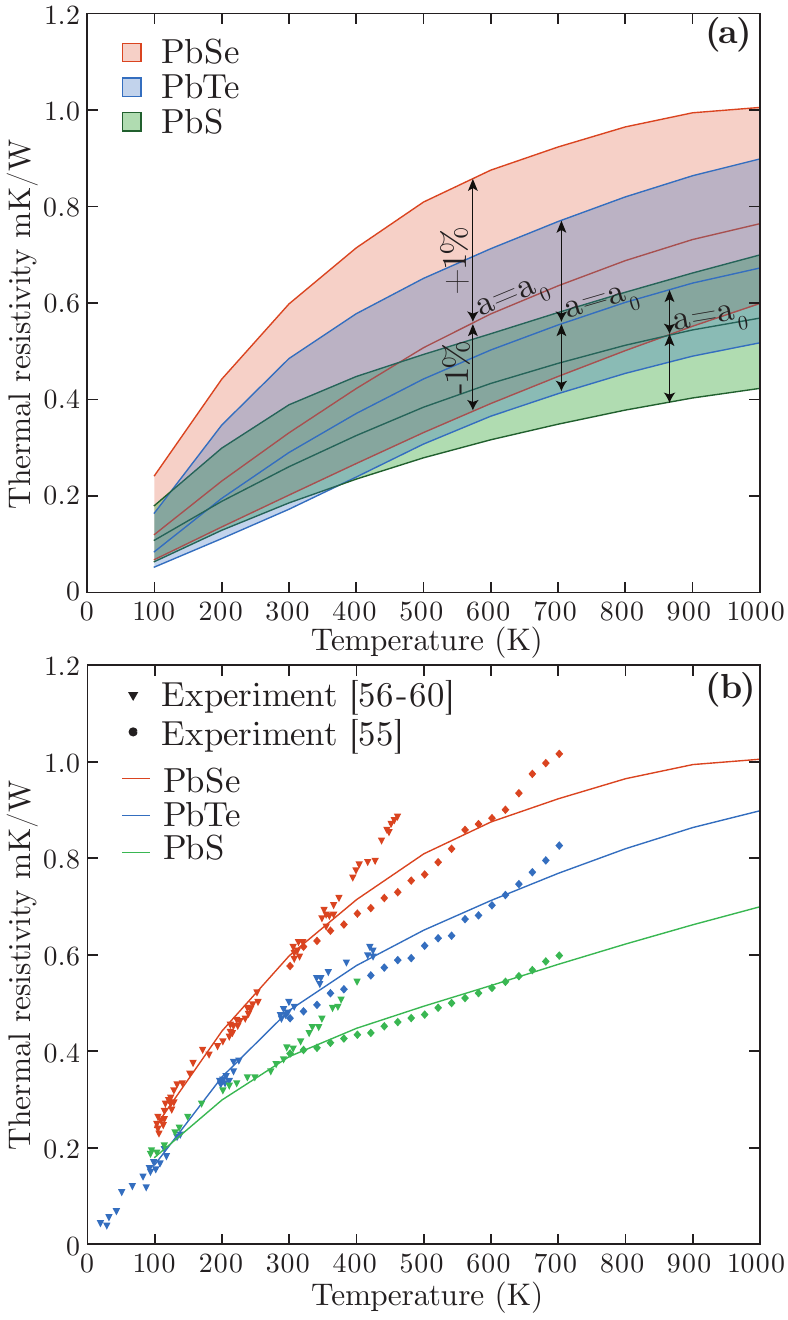}
\caption{\label{fig:tcond}  (Color online) a) Thermal resistivity of PbS, PbSe and PbTe as a function of temperature with $\pm$1\% variation of the lattice parameter. b) Thermal resistivity of PbS, PbSe and PbTe (lines) for +1\% lattice constant. Experimental data (symbols) for PbS, PbSe and PbTe are taken from the Ref.\cite{ElSharkawy1983} between 300 and 700 K. Experimental data at low temperatures for PbS are taken from Ref.\cite{greig_1960,ioffe_1960,ravich_1958}; for PbSe from Ref. \cite{greig_1960,Devyatkova1960} and for PbTe from Ref. \cite{greig_1960,Devyatkova1962}.}
\end{figure}

We first calculated the thermal resistivity, or the inverse of thermal conductivity, of PbS, PbSe and PbTe. Intuitively, the compound with the lightest element and highest frequencies, PbS, would be expected to have the highest thermal conductivity, while PbTe would be expected to have the lowest as the softest and heaviest of the compounds. However, experimentally PbSe has the lowest thermal conductivity, a feature that previous computational studies have failed to reproduce.

The lattice thermal resistivity for three compounds calculated with TDEP is plotted as a function of temperature in \figref{fig:tcond}. We first note that the thermal conductivity for three compounds is very sensitive to the volume as was previously found for PbTe~\cite{rabe_1985,an2008,Romero2015}. We plot thermal resistivity with $\pm 1\%$ variation of the lattice parameter obtained from DFT in \figref{fig:tcond}a), observing a factor of two variation in thermal resistivity. This factor is an intrinsic uncertainty in the DFT calculations. 

\figreffull{fig:tcond}b) shows the calculated thermal resistivity versus temperature along with experimental data. Our calculations are in good agreement with experimental data between 100 and 600 K \cite{Devyatkova1962,ElSharkawy1983} at volumes corresponding to +1\% increase of lattice parameter (\figref{fig:tcond}b)), which corresponds to the experimental lattice parameter.\cite{} Prior works have also found that a small modification of the lattice parameter was necessary to match experimental data \cite{lindsay2013,Romero2015}. In any case, the trends are unaffected by the choice of lattice parameter, and so the following analysis will be performed for the +1\% case. 

We observe that our calculation is able to reproduce two key trends. Firstly, we reproduce the strongly nonlinear behavior of the thermal resistivity with temperature. The change in slope of thermal resistivity around 350 K is unattainable with computational methods that derive quantities from 0 K calculations. The origin of the kink in thermal resistivity has already been explained as a decrease in scattering phase space with temperature due to the stiffening of TO mode \cite{Romero2015}. Secondly, our calculation correctly predicts the high thermal resistivity of PbSe in comparison to PbS and PbTe, which was not predicted by previous studies~\cite{tian_2012,Lee2014,Skelton2014}. This trend is unexpected because PbTe is heavier and softer than PbSe yet its thermal resistivity is consistently lower than that of PbSe over the entire temperature range.
\begin{figure*}
\includegraphics[width=\linewidth]{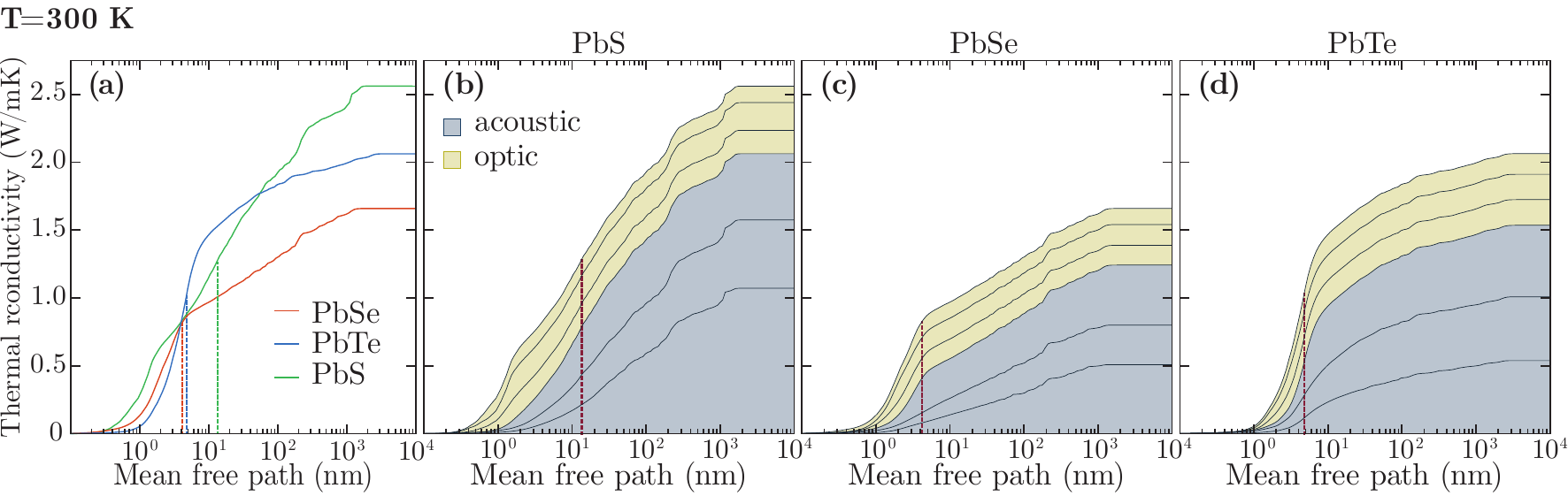}
\caption{\label{fig:mfps} (Color online) (a) Cumulative thermal conductivity as a function of mean free path at T=300 K for PbS, PbSe and PbTe. The vertical dashed lines indicates where 50$\%$ of thermal conductivity is accumulated. Cumulative thermal conductivity for (b) PbS, (c) PbSe, and (d) PbTe decomposed per branch. The grey region gives the contribution from the acoustic branches while the yellow region gives the contribution from the optical modes. In PbSe and PbTe 50\% of contribution to the thermal conductivity comes from the phonons with mean free paths smaller than 5 nm.}
\end{figure*}

To gain more insight into the phonon transport properties, we calculate the cumulative thermal conductivity versus mean free path at T=300 K.  \figreffull{fig:mfps}a) shows cumulative thermal conductivity as a function of mean free path for PbS, PbSe and PbTe. In both PbSe and PbTe 50\% of contribution to the thermal conductivity comes from phonons with mean free paths smaller than 4-5 nm. Further, we analyze the contributions to the total thermal conductivity accumulated from each branch for each compound as shown in \figrefmany{fig:mfps}b), c) and d). In all three cases, optical modes contribute a significant portion (52\%, 46\% and 40\% in PbSe, PbTe and PbS, respectively) to the thermal conductivity. Although the cumulative thermal conductivity distributions provide useful insights, the explanation of why the thermal conductivity of PbSe is lower than in PbTe remains unclear. 

\subsection{Spectral function $S(\mathbf{q},E)$}
\begin{figure*}
\includegraphics[width=\linewidth]{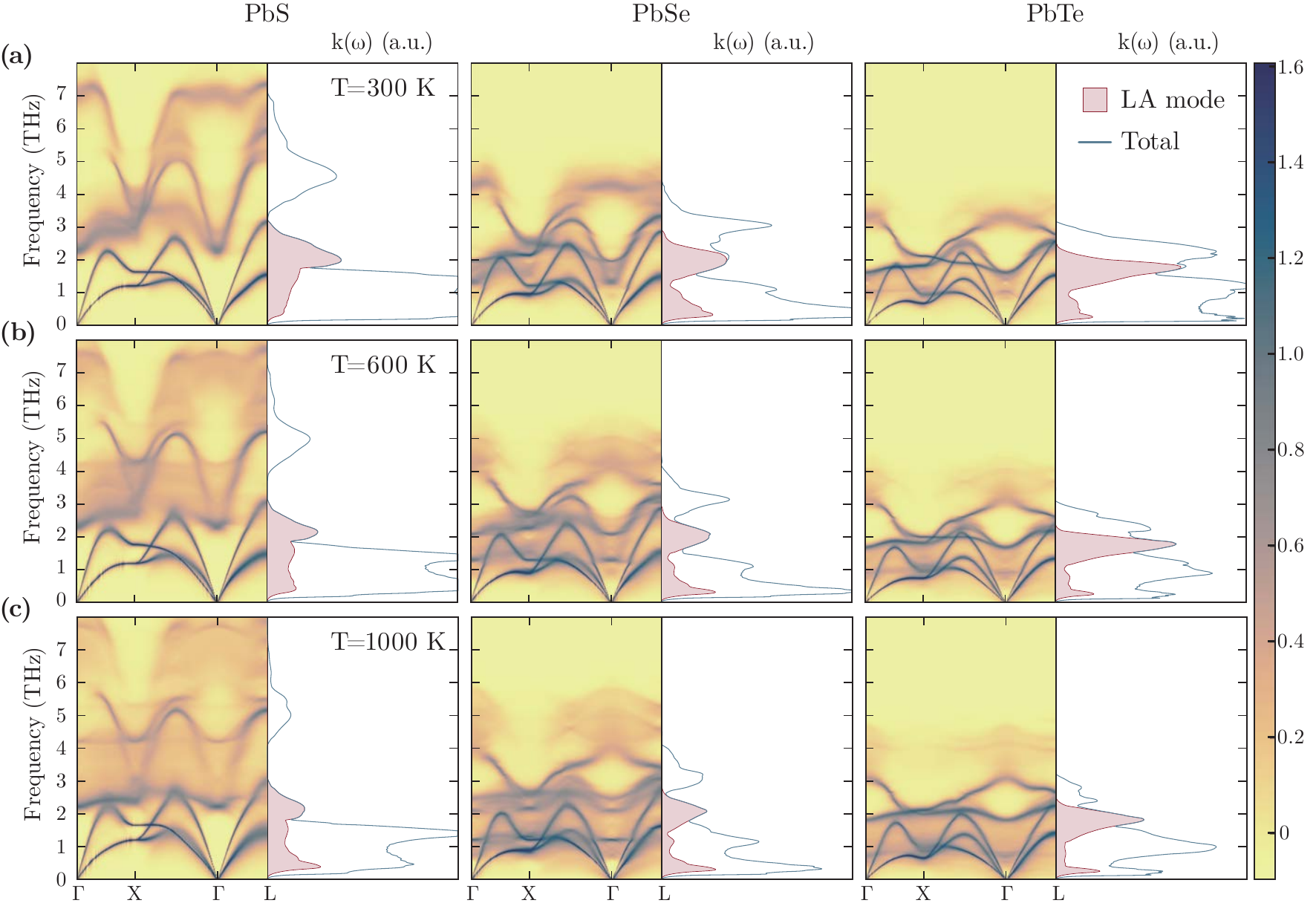}
\caption{\label{fig:lineshapes} (Color online) Left panels: spectral function (logarithmic intensity scale) along the high symmetry directions for PbS (left), PbSe (middle) and PbTe (right) at  (a) 300 K, (b) 600 K and (c) 1000 K. Right panels: total spectral thermal conductivity and spectral thermal conductivity for the longitudinal acoustic mode versus frequency.}
\end{figure*}

To proceed, we next calculate the spectral function $S(\mathbf{q},E)$, which describes the spectrum of lattice excitations with energy $E=\hbar\Omega$ that are not necessarily independent plane-waves. Starting with many-body perturbation theory, we calculate the frequency-dependent self-energy~\cite{maradudin_1962, Cowley1968} $\Sigma(\Omega) = \Delta(\Omega) + i\Gamma(\Omega)$, where the imaginary ($\Gamma(\Omega)$) part is
\begin{equation}\label{eq:gammaqs}
\begin{split}
\Gamma_{\vec{q}s}(\Omega)=& \sum_{s's''}
\frac{\hbar \pi}{16}
\frac{V}{(2\pi)^3}
 \iint_{\mathrm{BZ}}
\left|\Psi^{\vec{q}\vec{q}'\vec{q}''}_{ss's''}\right|^2 
\Delta_{\vec{q}\vec{q}'\vec{q}''}
\times \\
& \bigl[(n_{\vec{q}'s'}+n_{\vec{q}''s''}+1)
\delta(\Omega-\omega_{\vec{q}'s'}-\omega_{\vec{q}''s''}) \\
+ & 2(n_{\vec{q}'s'}-n_{\vec{q}''s''})
\delta(\Omega-\omega_{\vec{q}'s'}+\omega_{\vec{q}''s''}) \bigr]d\vec{q}'d\vec{q}'',
\end{split}
\end{equation}
and the real part is obtained via a Kramers-Kronig transformation of the imaginary part:
\begin{equation}\label{eq:gammareal}
\Delta(\Omega)=\frac{1}{\pi}\int \frac{\Gamma(\omega)}{\omega-\Omega}d\omega.
\end{equation}
The imaginary part of the self energy contains a sum over all possible three-phonon interactions between mode $s$ and $s's''$. $n_{\vec{q}s}$ are the Bose-Einstein occupation numbers for phonons with frequency $\omega_{\vec{q}s}$ at wave vector $\vec{q}$. The delta functions in Eq. \eqref{eq:gammaqs} ensure that energy and momentum are conserved. $\Psi^{\vec{q}\vec{q}'\vec{q}''}_{ss's''}$ are the three-phonon matrix elements.

From the self-energy we get the spectrum of possible excitations at energy $E$:
\begin{equation}\label{eq:lineshape}
S(\vec{q},E)
\propto
\sum_s
\frac{2\omega_{\vec{q}s}\Gamma_{\vec{q}s}(\Omega)}
{\left(\Omega^2-\omega^2_{\vec{q}s}-2\omega_{\vec{q}s}\Delta_{\vec{q}s}(\Omega)\right)^2
+4\omega^2_{\vec{q}s}\Gamma^2_{\vec{q}s}(\Omega)}
\end{equation}
This spectral function, or phonon lineshape, is shown in \figref{fig:lineshapes}. For the $S(\mathbf{q},E)$ calculations we used a 35$\times$35$\times$35 q-grid consistent with the thermal conductivity calculations. The tetrahedron method was used for numerical evaluation of the self-energy in Eq.~\eqref{eq:gammaqs}. The $S(\mathbf{q},E)$ of PbS is typical of a weakly anharmonic solid with Lorentzian broadening of single peaks. PbTe is more anharmonic than PbS and our calculation successfully reproduces the double peak structure observed previously~\cite{Delaire2011,li2014}. 

The $S(\mathbf{q},E)$ of PbSe, however, is quite unusual. We observe the formation of a dispersionless optical mode in the acoustic phonon frequencies as well as a kink in the dispersion of the LA branch. This mode has signatures of an intrinsic localized mode (ILM), also known as a discrete breather, that occurs due to strong anharmonicity \cite{sievers_1988,Flach_2008}. ILMs have been previously experimentally observed in NaI in the acoustic-optical phonon gap\cite{Manley2009}. In PbSe, the mode appears in the acoustic frequencies and reflects extremely large anharmonic scattering of the LA branch. The effect of this strong anharmonic interactions can be observed in the strong decrease in spectral thermal conductivity for the LA mode as in \figref{fig:lineshapes}. It is interesting to note that considerable focus has been placed on a similar anharmonic interaction in PbTe \cite{Delaire2011,Jiang2016} yet it does not exhibit an ILM.

\subsection{Lineshape at $\Gamma$}
\begin{figure}
\includegraphics[width=.5\linewidth]{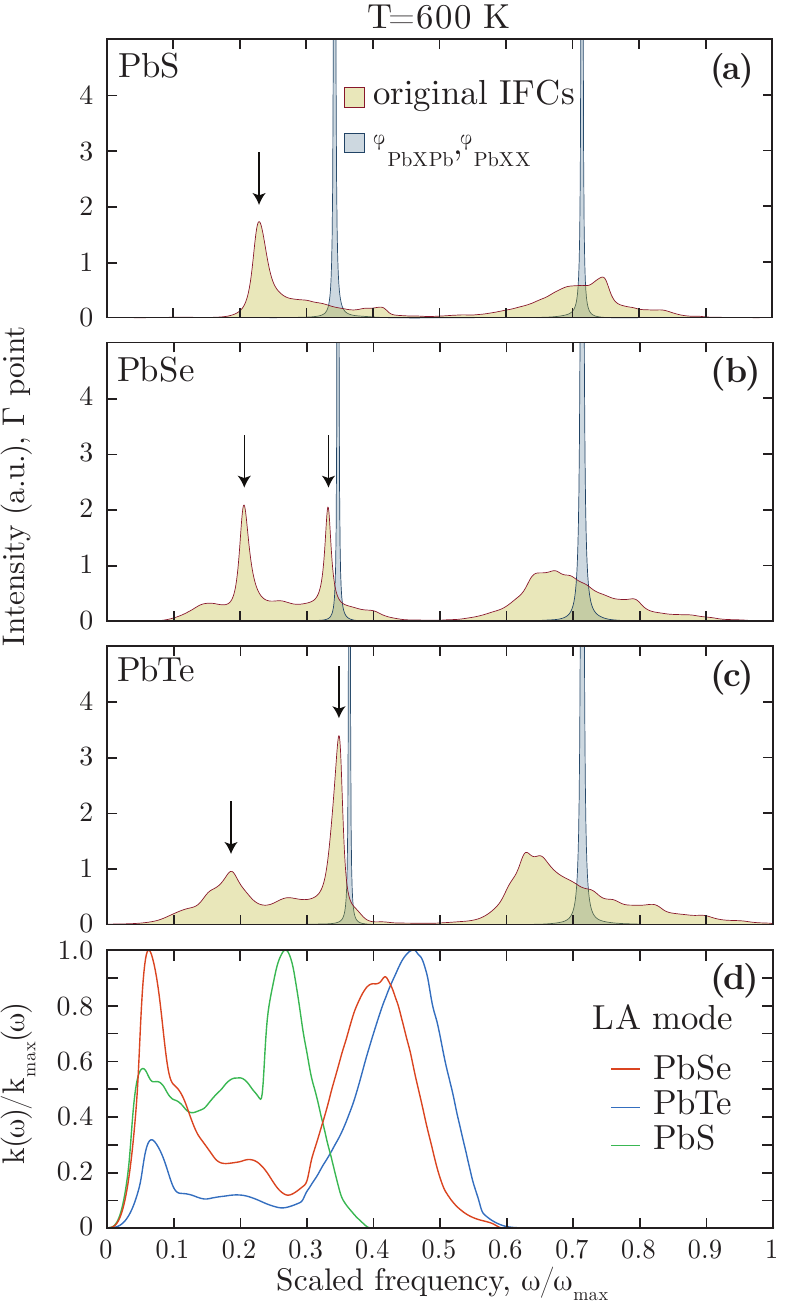}
\caption{\label{fig:lineshapes_gamma} (Color online) Phonon lineshapes for (a) PbS, (b) PbSe and (c) PbTe at 600 K at $\Gamma$. The arrows indicate the main and the secondary peaks with the original force constants. By setting the strongest three-body interaction to zero, in all cases the lineshapes revert to the narrow Lorentzian peaks. Comparing the lineshapes between PbSe and PbTe, we observe the split peak, but the secondary peaks differ. In PbTe the peak is broader and weaker than the main peak, but in PbSe it is much sharper and with equal intensity to the main peak, corresponding to an intrinsic localized mode. In (d) we show the impact of the secondary peak on scaled spectral thermal conductivity for the longitudinal mode. The dip in spectral thermal conductivity in PbSe is three times larger in magnitude than in PbTe, which is $\approx$~15\% and 5\% of the maximum spectral thermal conductivity for the LA mode, respectively.}
\end{figure}

To understand the origin of the ILM, we analyze the cubic IFCs responsible for the strength of the thee phonon interactions. We have considered three-body interactions within the first six coordination shells, and sequentially set the irreducible IFCs to zero while recalculating the lineshape at the $\Gamma$-point. We identified two force constants, corresponding to the nearest neighbor cubic interactions of degenerate triplets (PbXPb or PbXX, where X is S, Se or Te) in the [100] direction, that strongly influence the lineshape. These two force constants are linked, since any displacement of a nearest neighbor Pb-X pair will involve both force constants, which have opposite signs and are related via the translational invariance condition. In \figrefmany{fig:lineshapes_gamma}a), b) and c) we show the phonon lineshapes at $\Gamma$ when both cubic IFCs are set to zero. The lineshapes become Lorentzian, typical of a weakly anharmonic solid, indicating that this interaction is responsible for the unusual lineshapes.

The double peak structure in the case of PbTe has already been reported in Refs. \cite{Delaire2011,li2014}. In PbSe we find the similar behaviour of the TO mode except that the anharmonic interaction is even stronger and results in a secondary peak, the ILM, with the same intensity as the first. In PbS we identify only the anharmonic broadening of the single peak. Removing the nearest neighbor cubic interactions corresponding to these force constants result in a factor of 10 or more increase in thermal conductivity, indicating that this force interaction is the dominant source of scattering. Importantly, in this case the thermal conductivity of PbSe is higher than PbTe, indicating that this interaction is the origin of low thermal conductivity of PbSe.

\figreffull{fig:lineshapes_gamma}d) shows the scaled spectral thermal conductivity for LA mode calculated with original force constants that is the most affected by interaction with TO mode in PbSe and PbTe. The secondary peak causes the dip in spectral thermal conductivity around 0.2-0.3 of the scaled frequency. In PbSe the dip is three times larger in magnitude than in PbTe, reflecting the strong interaction in PbSe. The dip is not observed in PbS, since the TO-LA interaction is weaker.

We can therefore conclude that the low thermal conductivity of PbSe is related to the anomalously large anharmonic interaction between the LA and TO branches. Interestingly, though, we have been unable to replicate the ILM and low thermal conductivity in either PbS or PbTe by swapping harmonic or cubic IFCs with those of PbSe. This observation indicates that the presence of the ILM and unusually low thermal conductivity of PbSe is not solely due to a specific interaction but rather the overall interplay of the harmonic and anharmonic force constants. The precise origin of the ILM will be the topic of a future work.

\subsection{Atomic pair distribution.}
\begin{figure*}
\includegraphics[width=\linewidth]{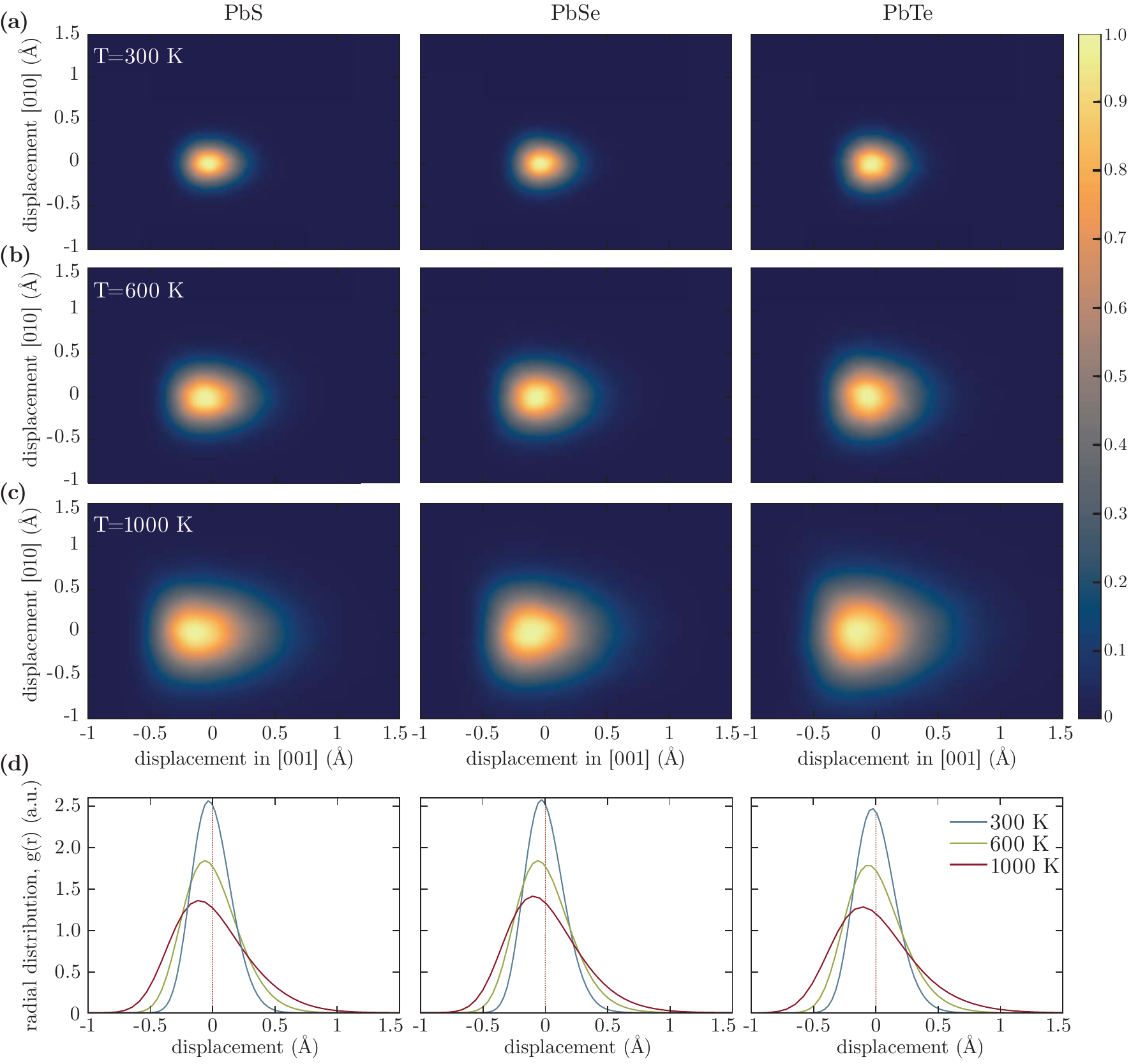}
\caption{\label{fig:atomicdistr} (Color online) Pair vector distribution of PbS (left), PbSe (middle) and PbTe (right) projected on the [001] and [010] planes at (a) 300 K, (b) 600 K and (c) 1000 K. Radial distribution function is shown in (d). There is a strong asymmetry of the peak around the equilibrium position, indicative of strong anharmonicity, however the centre of mass is positioned exactly at the zero and no off-centering observed. In a harmonic material these distributions would be Gaussian.}
\end{figure*}

Finally, we confirm the strong cubic nearest neighbor force constants by calculating the radial distribution function, also known as the pair correlation function, defined as,
\begin{equation}
g(r) = \frac{ n(r) }{\rho 4 \pi r^2 dr}\,,
\end{equation}
where $\rho$ is the mean particle density, and $n(r)$ the number of particles in an infinitesimal shell of width $dr$. Usually, this quantity is averaged over all atoms in the system. Here we project $g(r)$ onto symmetrically equivalent pairs, giving a projected pair distribution:
\begin{equation}
g_i(r) = {\rho 4 \pi r^2 dr} \sum_i \delta\left( \left|r_i\right|-r \right)\,,
\end{equation}
where the index $i$ corresponds to a coordination shell. The coordination shell is defined from the ideal lattice as the set of pairs that can transform to each other via a space group operation. In addition we also calculated the symmetry-resolved histograms of pair vectors, a histogram of all the pair vectors accumulated over time from Born-Oppenheimer molecular dynamics. The simulations were carried using Born-Oppenheimer molecular dynamics with thermalized configurations as a starting point at equilibrium volumes at 300, 600 and 1000 K for 22 ps with a time step of 2 fs with the same settings as discussed in the \secref{sec:methods}. The temperature was controlled using a Nos\'e thermostat \cite{vasp_nose1984}. 

The radial distribution function and pair vector distributions for the first coordination shell are shown in \figrefmany{fig:atomicdistr}a), b) and c). The displacements become more asymmetric with temperature increase. The asymmetry is clearly seen when the distributions are integrated to the projected pair distribution functions in \figref{fig:atomicdistr}d). The strong asymmetry of the peak only proves that the displacements of these materials are affected by anharmonic force constants. Our calculation clearly shows that the center of mass of the distributions is exactly at the equilibrium pair distance. We conclude, in line with \citet{Keiber2013}, that Pb is not off-center in PbSe, similarly to the case of PbTe. 

\section{Conclusions}

We used TDEP with an efficient scheme to generate stochastic thermalized configurations to investigate the thermal properties of PbS, PbSe and PbTe, and particularly the unusually low thermal conductivity of PbSe. Our calculation successfully reproduces the nonlinear thermal resistivity with temperature trend as well as the low thermal conductivity of PbSe, in contrast to prior \emph{ab-initio} calculations. By computing the phonon spectral function, we identified an intrinsic localized mode in PbSe in the acoustic frequencies that reflects an extremely strong anharmonic interaction between the LA and TO branches. Our work shows the deep insights into thermal phonons that can be obtained from \emph{ab-initio} calculations that are not confined to the weak limit of anharmonicity. 

\section{Acknowledgments}

N.S. and A.J.M. acknowledge the support of the DARPA MATRIX program under grant number HR0011-15-2-0039. O.H. acknowledges the support from the Swedish Research Council (VR) program 637-2013-7296. This work used the Extreme Science and Engineering Discovery Environment (XSEDE), which is supported by National Science Foundation grant number ACI-1053575 and the Swedish National Infrastructure for Computing (SNIC) at PDC center (High Performance Computing at the KTH Royal Institute of Technology) and NSC center (Link\"oping University).

\end{document}